\documentclass[preprint,showpacs,aps]{revtex4}
\begin{document}
\def\be{\begin{equation}}
\def\ee{\end{equation}}
\def\bea{\begin{eqnarray}}
\def\eea{\end{eqnarray}}
\title{Unified geometric description of black hole thermodynamics}
\author{Jos\'e L. \'Alvarez}
\email{jose.alvarez@nucleares.unam.mx}
\affiliation{ Instituto de Ciencias Nucleares\\
Universidad Nacional Aut\'onoma de M\'exico  \\
A.P. 70-543   \\
04510 M\'exico D.F., MEXICO}
\author{Hernando Quevedo}
\email{quevedo@nucleares.unam.mx}
\affiliation{ Instituto de Ciencias Nucleares\\
Universidad Nacional Aut\'onoma de M\'exico  \\
A.P. 70-543   \\
04510 M\'exico D.F., MEXICO}
\author{Alberto S\'anchez}
\email{asanchez@nucleares.unam.mx}
\affiliation{ Instituto de Ciencias Nucleares\\
Universidad Nacional Aut\'onoma de M\'exico  \\
A.P. 70-543   \\
04510 M\'exico D.F., MEXICO}
\begin{abstract}
In the space of thermodynamic equilibrium states  
we introduce a Legendre invariant metric which contains all
the information about the thermodynamics of black holes. 
The curvature of this thermodynamic metric becomes singular at those points
where, according to the analysis of the heat capacities, 
phase transitions occur. This result is valid for the Kerr-Newman
black hole and all its special cases and, therefore, provides
a unified description of black hole 
phase transitions in terms of curvature
singularities.  

\end{abstract}

\maketitle

\section{Introduction}
\label{sec:int}

According to the no-hair theorems of Einstein-Maxwell theory, 
electro-vacuum black holes are completely described by three
parameters only: mass $M$, angular momentum $J$, and electric charge 
$Q$. The corresponding gravitational field is described by the 
Kerr-Newman metric which in Boyer-Lindquist coordinates can be 
expressed as \cite{solutions}
\bea
ds^2 = &-&\frac{\Delta - a^2\sin^2\theta}{\Sigma} dt^2 
-\frac{2a\sin^2\theta (r^2+a^2 -\Delta)}{\Sigma} dtd \varphi \nonumber\\
& +& \frac{(r^2+a^2)^2 - a^2\sin^2\theta\, \Delta }{\Sigma} \sin^2\theta d\varphi^2
+\frac{\Sigma}{\Delta} dr^2 + \Sigma d\theta^2 \ , 
\label{kn}
\eea
\be 
\Sigma = r^2 + a^2\cos^2\theta\ , \quad \Delta = (r-r_+)(r-r_-)\ ,
 \quad r_\pm = M \pm \sqrt{M^2-a^2-Q^2} \ ,
 \label{horkn}
\ee
where $ a = J/M$ is the specific angular momentum. 
Bekenstein \cite{bek73} discovered in 1973 that the horizon area $A$ of 
a black hole behaves as the entropy $S$ of a classical thermodynamic system. 
This was the beginning of what is now called thermodynamics
of black holes \cite{bch73,haw75,davies}. Although its statistical origin 
is still very unclear, black hole thermodynamics has been the subject of 
intensive research for the past three decades, 
due in part to its possible connection to a hypothetical 
theory of quantum gravity. 

It has been established that the physical parameters of the Kerr-Newman black hole 
satisfy the first law of black hole 
thermodynamics \cite{bch73}
\be 
dM = T dS +  \phi d Q +\Omega_H  d J  \ ,
\label{first}
\ee
where  $T$ is the Hawking temperature which is proportional to the surface 
gravity on the horizon, $S=A/4$ is the entropy,
 $\Omega_H $ is the angular velocity on the horizon, 
and $\phi$ is the electric potential. As in ordinary thermodynamics, all the 
thermodynamic information is contained in the fundamental equation which was 
first derived by Smarr \cite{smarr73} 
\be
M = \left[ \frac{\pi J^2}{S} + \frac{S}{4\pi}\left(1 + \frac{\pi Q^2}{S}\right)^2\right]^{1/2}\ .
\label{feqbh}
\ee 
In the entropy representation, this fundamental equation can be rewritten as
\be
S= \pi\left( 2 M^2 - Q^2 +2\sqrt{M^4-M^2Q^2 - J^2}\right) \ .
\label{entropy}
\ee 
Davies \cite{davies} argued 
that black holes undergo a second order phase transition at the points where the heat
capacity diverges. This argument is supported by the result that some critical exponents
related to the singular points obey scaling laws \cite{sokmaz,lau1,lau2,munpir,lou1,lou2}.
Following Davies, we assume in this work that the structure of the phase transitions 
of the Kerr-Newman black hole is determined by the corresponding heat capacity 
$C=T(\partial S/\partial T)$: 
\be
C_{Q,J} = -\frac{4TM^3S^3}
{2M^6-3M^4Q^2-6M^2J^2+Q^2J^2 + 2(M^4-M^2Q^2 - J^2)^{3/2} } \ .
\label{capkn}
\ee

On the other hand, differential geometric concepts have been applied in 
ordinary thermodynamics since the seventies. First, Weinhold \cite{wei1}
introduced on the space of equilibrium states a metric whose components are given as 
the Hessian of the internal thermodynamic energy.  
Later, Ruppeiner \cite{rup79} introduced a metric which is defined as minus 
the Hessian of the entropy, and is conformally equivalent to Weinhold's metric,
with the inverse of the temperature as the conformal factor.
One of the aims of the application of geometry in thermodynamics is to 
describe phase transitions in terms of curvature singularities so that the
curvature can be interpreted as a measure of thermodynamic interaction. 
This turns out
to be true in the case of the ideal gas, whose curvature vanishes, and the
van der Waals gas for which the curvature of 
Weinhold's and Ruppeiner's metric becomes singular at those points where 
phase transitions occur. This is an encouraging result that illustrates  
the applicability of geometry in thermodynamics. It is then natural to 
try describe the phase transitions of black holes in terms of curvature
singularities in the space of equilibrium states. 
Unfortunately, the obtained results are contradictory. 
For instance, for the Reissner-Nordstr\"om black hole the Ruppeiner metric is flat
\cite{aman03},
whereas the Weinhold metric with the mass as thermodynamic potential 
presents a curvature singularity only in the limit of an extremal black hole. 
None of these results reproduces the phase transitions as predicted by Davies
using the heat capacity. Nevertheless, a simple change of the thermodynamic potential 
\cite{shen05} affects Ruppeiner's geometry in such a way  that the resulting curvature
singularity now corresponds to a phase transition. A dimensional reduction 
of Ruppeiner's curvature seems to affect its properties too \cite{mza07}. 
The situation is similar in the
case of the Kerr black hole: Weinhold's metric is flat \cite{aman03}, and the original 
Ruppeiner metric does not present curvature singularities at the points of phase 
transitions of the Kerr black hole. Nevertheless, with a change of thermodynamic potential 
\cite{shen05}, Ruppeiner's metric reproduces the structure of the phase transitions of
the Kerr black hole. These results seem to indicate that, in the case of black holes,
geometry and thermodynamics are compatible only for a very specific thermodynamic 
potential. However, it is well known that ordinary thermodynamics does not depend on 
the thermodynamic potential. We believe that a geometric description of thermodynamics
should preserve this property, i.e., it should be invariant with respect to 
Legendre transformations.

Recently \cite{quev07}, the formalism of geometrothermodynamics (GTD) 
was proposed as a geometric approach that incorporates Legendre invariance
in a natural way, and allows us to derive Legendre invariant 
metrics in the space of equilibrium states. Since Weinhold and Ruppeiner metrics
are not Legendre invariant, one of the first results in the context of GTD was 
the derivation of simple Legendre invariant generalizations of these metrics and their
application to black hole thermodynamics. It 
turned out \cite{quev08} that the thermodynamics of the Reissner-Nordstr\"om black hole
is compatible with both Weinhold and Ruppeiner generalized metric structures. However,
in the case of the Kerr black hole both generalized geometries are flat and, therefore,
cannot reproduce its thermodynamic behavior. This was considered as a negative result 
for the use of geometry in black hole thermodynamics.

In the present work we use GTD to derive a Legendre invariant metric which 
completely and consistently reproduces the thermodynamic behavior of black 
holes, including the Kerr-Newman black hole. This result finishes the controversy 
regarding the application of geometric structures in black hole thermodynamics.
The phase transition structure contained in the heat capacity of black holes
becomes completely integrated in the scalar curvature of the Legendre invariant
metric so that a curvature singularity corresponds to a phase transition. 


This paper is organized as follows. In section \ref{sec:gtdbh} we introduce 
the general formalism of GTD for black holes.
A particular Legendre invariant metric is given in the thermodynamic phase 
space which is the starting point of our analysis. In section  \ref{sec:gtd2} 
 we apply 2-dimensional GTD 
in its entropy representation to the Reissner-Nordstr\"om
and Kerr black holes. The analysis of the Kerr-Newman black hole 
requires 3-dimensional GTD and it is presented in section \ref{sec:gtd3}.
 Finally, section \ref{sec:con} is devoted
to discussions of our results and suggestions for further research.
Throughout this paper we use units in which $G=c=k_{_B}=\hbar =1$.

\section{Geometrothermodynamics of black holes }
\label{sec:gtdbh}

The starting point of GTD is the thermodynamic phase space ${\cal T}$
which in the case of Einstein-Maxwell black holes can be defined as
a 7-dimensional space with coordinates 
$Z^A=\{M,S,Q,J,T,\phi,\Omega_H\}$, $A=0,...,6$.  
In the cotangent space ${\cal T}^*$, we introduce the fundamental one-form
\be
\Theta_M = dM - T dS - \phi dQ - \Omega_H d J \ ,
\ee
which satisfies the condition $\Theta_M \wedge (d\Theta_M)^3 \neq 0$. Furthermore,
in ${\cal T}$ we introduce a non degenerate metric $G$. The triplet 
$({\cal T},\Theta_M,G)$ is said to form a Riemannian contact manifold.
Let ${\cal E}$ be a 3-dimensional subspace of ${\cal T}$ with coordinates 
$E^a=\{S,Q,J\}$, $a=1,2,3$, defined by means of a smooth mapping 
$\varphi_M: {\cal E} \longrightarrow {\cal T}$. The subspace ${\cal E}$ 
is called the space of equilibrium states if $\varphi_M^*(\Theta_M)=0$,
where $\varphi_M^*$ is the pullback induced by $\varphi_M$. Furthermore, 
a metric structure
$g$ is naturally induced on ${\cal E}$ by applying the pullback on the metric 
$G$ of ${\cal T}$, i.e., $g=\varphi_M^*(G)$. It is clear that the condition 
$\varphi_M^*(\Theta_M)=0$ leads immediately to the first law of thermodynamics of
black holes as given in Eq.(\ref{first}). It also implies the existence
of the fundamental equation $M=M(S,Q,J)$ and the conditions of thermodynamic 
equilibrium
\be
T =\frac{\partial M}{\partial S} \ , \quad
\phi =\frac{\partial M}{\partial Q} \ ,\quad
\Omega_H  =\frac{\partial M}{\partial J} \ .
\ee

Legendre invariance is an important ingredient of GTD. It allows us to change 
the thermodynamic potential without affecting the results. If we denote the
intensive thermodynamic variables as $I^a = \{T,\phi,\Omega_H\}$, then a Legendre
transformation is defined by
 \cite{arnold}
\be
\{M, E^a,I^a\}\longrightarrow \{\tilde M, \tilde E ^a, \tilde I ^ a\}
\ee
\be
 M = \tilde M - \delta_{ab} \tilde E ^a \tilde I ^b \ ,\quad
 E^a = - \tilde I ^ {a}, \ \  
 I^{a} = \tilde E ^ a \ .
 \label{leg}
\ee
It is easy to see that the fundamental one-form $\Theta_M$ is invariant with respect 
to Legendre transformations. Furthermore, if we demand that the metric $G$ be Legendre
invariant, it can be shown \cite{quev07} that the induced metric $g=\varphi_M^*(G)$ is
also Legendre invariant. 

Another advantage of the use of GTD is that it allows us to easily implement different 
thermodynamic representations. The above description is called the $M-$representation because
the fundamental equation is given as $M=M(S,Q,J)$. However, one can rewrite this
equation as $S=S(M,Q,J)$, $Q=Q(S,M,J)$ or $J=J(S,M,Q)$, and redefine the coordinates
in ${\cal T}$ and the smooth mapping $\varphi$ in such a  way that the condition 
$\varphi^*(\Theta)=0$ generates on ${\cal E}$ the corresponding fundamental equation
in the $S-$, $Q-$, or the $J-$representation, respectively. As an example of this 
procedure we will present the $S-$representation which turned out to be the most appropriate 
for the description of black hole thermodynamics.  It must be emphasized, however, 
that the results obtained with different representations of the same
fundamental equation are completely equivalent.  

For the $S-$representation we consider the fundamental one-form 
\be
\Theta_S =  dS -\frac{1}{T} dM  +\frac{ \phi}{T} d Q + \frac {\Omega_H }{T} d J \  ,
\ee
so that the coordinates of ${\cal T}$ are $Z^A = \{S, E^a, I^a\}=\{S, M, Q, J, 1/T, 
-\phi/T,-\Omega_H/T\}.$
The space of equilibrium states ${\cal E}$ can then be introduced with 
the smooth mapping
\be
\varphi_S: \{M,Q,J\} \longmapsto \left\{ M, S(M,Q,J),Q,J, I^a(M,Q,J)\right\} \ ,
\ee
which, from the condition $\varphi^*_S(\Theta_S) =0$,
generates the first law of thermodynamics of black holes (\ref{first}) 
and the equilibrium conditions 
\be
\frac{1}{T} = \frac{\partial  S}{\partial M} \ ,\quad
\frac{\phi}{T} = -\frac{\partial  S}{\partial Q} \,\quad
\frac{\Omega_H }{T} = -\frac{\partial  S}{\partial J} \ . 
\ee
In this representation the fundamental equation is given as in Eq.(\ref{entropy}).

Consider now the following  metric on ${\cal T}$
\bea
G=  & &\left(dS - \frac{1}{T} dM  +\frac{ \phi}{T} d Q + \frac {\Omega_H }{T} d J \right)^2
 \nonumber \\
& & +\left(\frac{M}{T}-\frac{Q\phi}{T} - \frac{J\Omega_H}{T}\right)
\left[ dM d\left(\frac{1}{T}\right) + dQ d\left(\frac{\phi}{T}\right) + 
d J d\left(\frac{\Omega_H}{T}\right)\right] \ .
\eea
It is easy to show that this metric is invariant with respect to Legendre transformations
(\ref{leg}). The first term of this metric can be written in the form 
$\Theta_S\otimes \Theta_S$ 
so that its projection on ${\cal E}$ vanishes, due to the condition 
$\varphi^*_S(\Theta_S)=0$. Nevertheless, this term is necessary in order for the 
metric $G$ to be non degenerate. For the metric induced on ${\cal E}$ by means
of $g= \varphi_S^*(G)$  only the second term of $G$ is relevant. A straightforward 
computation leads to 
\be
g= \left(MS_M +Q S_Q + J S_J\right)\left( S_{MM} d M^2 - S_{QQ} dQ^2 - S_{JJ} dJ^2
-2 S_{QJ} dQ dJ\right) \ ,
\label{ge}
\ee
where for simplicity we introduced the notation that 
a subindex represents partial derivative with respect to 
the corresponding coordinate. This metric is Legendre invariant and non degenerate
and therefore can be used to introduce a Legendre invariant, Riemannian metric 
structure in the space of equilibrium states ${\cal E}$. This turns ${\cal E}$
into a well-defined Riemannian submanifold of the thermodynamic phase space ${\cal T}$. 
In the next sections we will show that metric (\ref{ge}) correctly reproduces 
the thermodynamic behavior of Einstein-Maxwell black holes.

\section{Black holes with two degrees of freedom}
\label{sec:gtd2}

From the above description of GTD, it follows that the dimension of the phase space 
is  $2n+1$, where $n$ is the number of thermodynamic degrees of freedom which 
coincides with the dimension of the subspace ${\cal E}$. The case $n=1$ 
corresponds to the Schwarzschild black hole with the mass $M$ as the only non vanishing
thermodynamic degree of freedom. In this case the Riemannian structure of ${\cal E}$ is
trivial. For $n=2$ the geometric structure of ${\cal E}$ is non trivial and corresponds to 
the Reissner-Nordstr\"om black hole $(J=0)$ or to the Kerr black hole $(Q=0)$. Notice 
that in this case the metric $g$ on ${\cal E}$ becomes diagonal what drastically simplifies
the calculations. 
The general Kerr-Newman black hole corresponds to a 3-dimensional 
manifold ${\cal E}$ with a non diagonal metric $g$. It requires a separate analysis that will
be performed in section \ref{sec:gtd3}.

\subsection{The Reissner-Nordstr\"om black hole}
\label{sec:rn}

The Reissner-Nordstr\"om metric can be obtained from Eq.(\ref{kn}) by imposing the condition 
$J=0$. It describes a static, spherically symmetric black hole with two horizons situated
at 
\be
r_\pm = M\pm\sqrt{M^2-Q^2}\ . 
\ee
We assume that $Q\leq M$ in order to avoid  
naked singularities. The thermodynamic information of this black hole is contained in 
the fundamental equation which, in the entropy representation we are using in this work, becomes
\be
S=\pi \left(M+\sqrt{M^2-Q^2}\right)^2 \ .
\label{srn}
\ee

According to Davies \cite{davies}, the phase transition structure of the Reissner-Nordstr\"om 
black hole can be derived from the heat capacity
\be
C_{Q} = \frac{4TM^3S^3}
{- 2M^6 + 3M^4 Q^2 - 2(M^4-M^2Q^2)^{3/2} } 
=-\frac{2\pi^2 r_+^2 (r_+ - r_-)}{r_+ - 3r_-} \ .
\label{caprn}
\ee
For our geometric approach to black hole thermodynamics all what is needed is the
fundamental equation as given in (\ref{srn}) from which we can calculate the 
thermodynamic metric 
\be
g^{RN}_{ab} = (MS_M + Q S_Q)\left(
\begin{array}{cc}
S_{MM}& 0  \\
0 & - S_{QQ} 	
\end{array}
\right)
= \frac{8\pi^2r_+^3}{(r_+-r_-)^3}\left(
\begin{array}{cc}
2r_+(r_+-3r_-) & 0 \\
0 & r_+^2 + 3r_-^2
\end{array}
\right)
 \ .
\label{grn}
\ee
Notice that this metric is singular in the extremal limit $r_+=r_-$. 
It could indicate a breakdown of our geometric approach. However, 
the analysis of the corresponding scalar curvature  
\be
R^{RN} = \frac{(r_+^2 - 3r_-r_++6r_-^2)(r_++3r_-)(r_+-r_-)^2}{
\pi^2 r_+^3 (r_+^2+3r_-^2)^2 (r_+-3r_-)^2}
\ee
shows that in the extremal limit the space of equilibrium states becomes flat.
This means that there must exist a different coordinate system in which the metric 
(\ref{grn}) does not diverges in the extremal limit. Moreover, we see from the
expression for the scalar curvature that the only singular point corresponds
to the value $r_+ = 3 r_-$ which is exactly the point where a phase transition 
occurs in the heat capacity (\ref{caprn}).

\subsection{The Kerr black hole}
\label{sec:kerr}

The Kerr metric corresponds to the limit $Q=0$ of the Kerr-Newman metric (\ref{kn}).
It describes the gravitational field of a stationary, axially symmetric, rotating
black hole with two horizons situated at the radial distances
\be
r_\pm = M \pm \sqrt{M^2 - J^2/M^2}\ .
\ee
The corresponding thermodynamic 
fundamental equation in the entropy representation becomes
\be
S=2\pi\left(M^2 + \sqrt{M^4-J^2}\right) \ .
\ee
Furthermore, second order phase transitions occur at the points where the 
heat capacity 
\be
C_{J} = \frac{4TM^3S^3}
{6M^2 J^2 - 2M^6 - 2(M^4-J^2)^{3/2} } 
=  \frac{2\pi^2 r_+ (r_++r_-)^2(r_+-r_-)}{ r_+^2-6r_+r_--3r_-^2}
\label{capkerr}
\ee
diverges. We assume values of the mass in the range $M^2\geq J$, the equality being 
the extremal limit of the Kerr black hole in which the two horizons coincide.

The Legendre invariant metric reduces in this case to 
\bea
g^{K}_{ab} & = & (MS_M + J S_J)\left(
\begin{array}{cc}
S_{MM}& 0  \\
0 & - S_{JJ} 	
\end{array}
\right) \nonumber \\
&=& \frac{16\pi^2r_+^2(r_++r_-)}{(r_+-r_-)^4}\left(
\begin{array}{cc}
r_+(r_+^2-6r_+r_--3r_-^2) & 0 \\
0 & r_+ + r_-
\end{array}
\right)
 \ .
\label{gkerr}
\eea
We obtain again a metric that becomes singular at the extremal limit $r_+=r_-$. 
The scalar curvature for the thermodynamic metric of the Kerr black hole can be 
expressed as 
\be
R^K = \frac{(3r_+^3+3r_+^2 r_- + 17 r_+r_-^2 + 9 r_-^3)(r_+-r_-)^3}
{2\pi^2 r_+^2(r_++r_-)^4(r_+^2-6r_+r_--3r_-^2)^2} \ .
\ee
This shows that the metric singularity at $r_+=r_-$ is only a coordinate singularity.
On the other hand, the curvature singularities are situated at the roots of the 
polynomial  equation $r_+^2-6r_+r_--3r_-^2=0$. According to the expression for the 
heat capacity (\ref{capkerr}), these are exactly the roots that determine the critical
points where phase transitions take place. 

\section{The general Kerr-Newman black hole}
\label{sec:gtd3}

The Kerr-Newman metric (\ref{kn}) describes the gravitational field of the
most general rotating, charged black hole. It possesses an outer horizon at $r_+$ and an
inner horizon at $r_-$, with $r_\pm$ given as in Eq.(\ref{horkn}). According 
to our results of section \ref{sec:gtdbh}, the space of thermodynamic 
equilibrium states is 3-dimensional and the corresponding Legendre invariant
metric can be written as
\be
g^{KN}_{ab}= (MS_M + QS_Q+ J S_J)\left(
\begin{array}{ccc}
S_{MM}& 0 & 0 \\
0 & - S_{QQ} & - S_{QJ} \\
0 & - S_{QJ} & - S_{JJ} 	
\end{array}
\right) \ .
\label{gkn1}
\ee
Inserting here 
the expression for the entropy (\ref{entropy}) we obtain a rather cumbersome
metric which cannot be written in a compact form. Moreover, 
the scalar curvature can be shown to have the form 
\be
R^{KN} = \frac{N}{D}\ , \quad
D= 4 (MS_M + QS_Q+ J S_J)^3 ( S_{QJ}^2 - S_{QQ}S_{JJ} ) ^3 S_{MM}^2
\ee
so that replacing the entropy formula we obtain 
\bea
D\propto & &  \left[ 2M^4 - 2M^2Q^2 - J^2 + (2M^2-Q^2)(M^4-M^2Q^2 - J^2)^{1/2}\right]^3 
\nonumber \\
& & \times \left[M^4 + (M^2-Q^2)(M^4-M^2Q^2 - J^2)^{1/2}\right]^3 \nonumber \\
&& \times \left[2M^6-3M^4Q^2-6M^2J^2+Q^2J^2 + 2(M^4-M^2Q^2 - J^2)^{3/2}\right]^2 \ .
\eea
The first two terms in squared brackets can be shown to be always positive in the 
range $M^4\geq M^2Q^2+J^2$, which is a condition that guarantees the non existence
of naked singularities. The third term in squared brackets is exactly the denominator
of the heat capacity (\ref{capkn}). This proves that the curvature 
singularities of the thermodynamic metric $g^{KN}$ are situated at those points 
where phase transitions can occur. Moreover, it can be shown that the curvature 
vanishes in the case of an extremal black hole, $M^4 = M^2Q^2+J^2$. This resembles 
the behavior of the curvature of the thermodynamic metrics of the Reissner-Nordstr\"om
and Kerr black holes presented in the last section.

\section{Discussion and conclusions}
\label{sec:con}

Using the formalism of GTD, in this work we derived a metric for the space of 
equilibrium states of black holes which reproduces  
the thermodynamic behavior of Reissner-Nordstr\"om, Kerr, and Kerr-Newman 
black holes. The thermodynamic metric is derived from a Legendre invariant metric which 
is introduced in the thermodynamic phase space. In contrast to other metrics used 
previously in the literature, the curvature singularities of our metric reproduce
in a unified manner the phase transitions of black holes, if we assume that 
phase transitions correspond to divergences of the heat capacity.  This result 
shows that the curvature of our thermodynamic metric can be used as a measure 
of thermodynamic interaction for black holes. 

For all black holes of Einstein-Maxwell theory, 
the space of equilibrium states, equipped with
our thermodynamic metric, becomes singular at those points where phase transitions 
occur, and it is flat in the limit of extreme black holes, i.e. when the two horizons
coincide. This indicates that our thermodynamic metric is well-defined in the 
region $M^4-M^2Q^2 -J^2\geq 0$, except at the phase transition points where it 
becomes singular. Outside this region, our thermodynamic metric is not well-defined
because the fundamental equation becomes complex and cannot be used to generate
the geometric Riemannian structure of the space of equilibrium states. This is an
indication that the thermodynamic description of black holes cannot be extended into 
the region of naked singularities. This is also an indication that classical 
thermodynamics cannot be used for black holes of the size of the Planck length,
which is the extremal limit of applicability one would expect for classical 
thermodynamics.

We assumed in this work Davies' formulation of phase transitions for black holes. 
However, the interpretation of divergences in specific heats as phase transitions 
is not definitely settled and is still a subject of debate 
\cite{cur81,pav91,kok93,ook93}. In fact, what is really needed is a microscopic 
description which would couple to the macroscopic thermodynamics of black  holes. 
However,
such a macroscopic description must be related to a theory of quantum gravity which 
is still far from being formulated in a consistent manner. In the meantime, we can
only use the intuitive interpretation of phase transitions as it is known in 
classical thermodynamics.

The thermodynamic metric we propose in this work is intuitively simple, it can be 
written in a compact form, and it satisfies the mathematical compatibility 
conditions of GTD. However, we do not have whatsoever any interpretation
of its components in terms of any physical theory. We believe that Ruppeiner's metric
is the only know thermodynamic metric with a specific physical interpretation 
in the context of thermodynamic fluctuation theory. It would be interesting
to investigate the stability of the metric derived in this work, especially 
the different scenarios available in black hole thermodynamics \cite{rupp07}.
  
The computer algebra system REDUCE 3.8 was used for most of the calculations 
reported in this work.

\section*{Acknowledgements} 

One of us (H.Q.) would like to thank G. Ruppeiner for helpful correspondence and 
literature hints. 
This work was supported in part by Conacyt, M\'exico, grant 48601.





\end{document}